\documentclass[aps,twocolumn,nofootinbib,superscriptaddress,preprintnumbers,pra,10pt]{revtex4-1}

\usepackage{float}
\usepackage{arydshln}
\usepackage{colortbl}
\usepackage{amsmath,amssymb}
\usepackage{dsfont} 
\usepackage{hyperref}
\usepackage{graphicx}
\usepackage{enumitem}
\usepackage{arydshln}
\usepackage{mathtools}
\usepackage{bbold}
\usepackage{soul}
\usepackage{slashed}
\usepackage{amsmath,bm}
\usepackage{yfonts}
\usepackage{lipsum}
\usepackage{slashed}
\usepackage{mathtools}
\usepackage{physics}

\makeatletter
\g@addto@macro\bfseries{\boldmath}
\makeatother

\newcommand{\be} {\begin{equation}}
\newcommand{\ee} {\end{equation}}
\newcommand{\bea} {\begin{eqnarray}}
\newcommand{\eea} {\end{eqnarray}}

\usepackage{color}

\usepackage{contour}

\def\eq#1{{Eq.~(\ref{#1})}}

\def\fig#1{{Fig.~\ref{#1}}}

\def\sect#1{{Sect.~\ref{#1}}}

\begin{document}

\preprint{ZU-TH-34/25}
\title{The third-generation-philic WIMP: an EFT analysis}

\author{Georgios Demetriou}
\affiliation{Institut f\"ur Theoretische Physik, ETH Z\"urich,  8093 Z\"urich, Switzerland }
\author{Gino Isidori}
\author{Gioacchino Piazza}
\author{Emanuelle Pinsard}
\affiliation{Physik-Institut, Universit\"at Zu\"rich, CH-8057 Z\"urich, Switzerland}

\begin{abstract}
We consider fermionic and scalar dark matter (DM) candidates that couple predominantly to third-generation Standard Model fermions, describing their interactions within an effective field theory framework. We show that current direct-detection constraints on these interactions are more than an order of magnitude weaker than those for flavor-universal couplings: effective scales in the few-TeV range remain allowed by existing data, leaving open the possibility of a connection between this type of new physics and a solution to the electroweak hierarchy problem.
Imposing the observed relic abundance from thermal freeze-out within the same effective theory,
a well-defined region for a fermionic DM candidate with mass in the 1–2 TeV range emerges.
Notably, this region will be fully probed by upcoming direct-detection experiments. Finally, we show that additional parameter space for both fermion and scalar cases can be recovered by going beyond the effective theory, through the introduction of a suitable vector mediator enabling resonant DM annihilation.
\end{abstract}

\maketitle
\allowdisplaybreaks

\section{Introduction} 
\label{sec:intro}

Despite its remarkable success in describing the fundamental forces and particles of nature, the Standard Model (SM) of particle physics is widely recognized as an effective theory, valid only up to a certain energy scale. Compelling empirical evidences, along with theoretical considerations, point to the existence of physics beyond the SM (BSM). One of the most significant empirical indications of BSM physics is the existence of dark matter (DM), a non-luminous form of matter that accounts for approximately 85\% of the total matter content of the universe~\cite{Sofue:2000jx,Bartelmann:1999yn,Mateo:1998wg}. 
On the theoretical front, a major unresolved issue within the SM is the instability of the Higgs mass term, usually referred to as the electroweak hierarchy problem. The WIMP (Weakly Interacting Massive Particle) paradigm offers a unified framework to address both of these problems. It postulates a DM candidate with a mass in the 100 GeV to few TeV range which, if coupled to SM particles with interaction strengths characteristic of the weak scale, naturally yields the observed DM relic abundance via thermal freeze-out.

The theoretical appealing WIMP scenario is facing growing tension due to the continued null results from direct-detection experiments. These experiments aim to measure the interactions between DM particles and atomic nuclei, providing a direct probe of DM couplings to light SM fermions. Stringent bounds from collaborations such as XENON1T~\cite{XENON:2022avm}, XENONnT~\cite{XENON:2023cxc}, and LUX-ZEPLIN~\cite{LZ:2024zvo} have ruled out DM–light quark interactions mediated by effective scales in the few TeV range (see, e.g., Ref.~\cite{Pospelov:2000bq,Kurylov:2003ra,Kopp:2009qt,Fan:2010gt,Hill:2011be,Fitzpatrick:2012ib,Fitzpatrick:2012ix,Cirigliano:2012pq,Menendez:2012tm,Hill:2013hoa,Klos:2013rwa,Anand:2013yka,Barello:2014uda,Catena:2014uqa,Hill:2014yxa,Hoferichter:2015ipa,DEramo:2016gos,Bishara:2016hek,Hoferichter:2016nvd,Brod:2017bsw,Bishara:2017nnn,Bishara:2017pfq,DEramo:2017zqw,Hoferichter:2018acd,Chen:2018uqz,Brod:2018ust,Hoferichter:2020osn,Brod:2021xbb,Aebischer:2022wnl,Cirelli:2013ufw}). In contrast, scenarios in which DM predominantly couples to third-generation SM fermions remain comparatively less constrained, offering a fertile ground for theoretical and phenomenological exploration.

Leaving aside the DM problem, the idea of new physics (NP) mainly coupled to the third generation has an interesting twofold appeal,
as emphasized in recent literature. 
First, a dominant NP coupling to the third generation 
is a natural feature of BSM models addressing the origin of the Yukawa couplings (see, e.g., Ref.~\cite{Barbieri:2011ci,Panico:2016ull,Bordone:2017bld,Greljo:2018tuh,Barbieri:2021wrc,Fuentes-Martin:2022xnb,Davighi:2023iks} and references therein).
Second, NP coupled predominantly to third-generation fermions allow for an effective scale as low as 1-2~TeV from the combination of collider, electroweak, and flavor constraints, to be contrasted with the several TeV bounds on flavor-universal 
interactions~\cite{Allwicher:2023shc}. Such lower scale of new physics allow to  minimise the tuning in the Higgs sector and  is also  potentially within the reach of current and near-future experiments. 

In this paper, we explore the impact of dark matter direct-detection constraints on scenarios where a hypothetical DM candidate (either fermion or scalar, in the typical WIMP mass range) couples predominantly to the SM fermions of the third generation. Explicit BSM frameworks of this type have been constructed, mainly in the context of models addressing the so-called $B$-meson anomalies~\cite{Andrianov:1998hx,Belanger:2015nma,Bauer:2015boy,Altmannshofer:2016jzy,Falkowski:2018dsl,Cerdeno:2019vpd,Trifinopoulos:2019lyo,Guadagnoli:2020tlx,Baker:2021llj}. Conversely, we address the problem from a bottom-up perspective, employing a generic Effective Field Theory (EFT) approach. We postulate generic contact interactions between a stable  DM candidate and third generation fermions  at some high scale, generating the effective couplings to light quarks
-- which are crucial for direct detection based on nuclear recoil -- via renormalization group evolution down to the low scale. 

The paper is organised as follows: in Section~\ref{sec:setup} we introduce the setup, defining the EFT at the high scale and estimating the radiatively induced nucleon-DM interaction. In Section~\ref{sec:DDbounds} we analyse current bounds on the EFT coefficients from 
direct-detection experiments. In Section~\ref{sec:relics} we analyse the constraint on the parameter space of the model imposing the observed relic abundance from thermal freeze-out, both in the EFT and in a simplified model allowing for resonant DM annihilation.
The results are summarised in the Conclusions.

\section{Setup} 
\label{sec:setup}
We extend the SM by a new  field, $\psi_{\rm DM}$, 
which is assumed to be singlet under the SM gauge group and charged under a $Z_2$
symmetry -- under which all SM fields are neutral -- that makes it stable. We also assume that the interactions between $\psi_{\rm DM}$ and SM fields occurs via heavy mediators with masses above the electroweak scale and above the DM mass. 
Integrating out the heavy mediators at the tree level, the interactions are  described by an effective local Lagrangian of the type 
\begin{equation}
\mathcal{L}^{(0)}_\text{SM-DM} = 
\sum_{d,i} \frac{C_i^{(d)}}{\Lambda^{d-4}} \mathcal{O}_i^{(d)} 
(\psi_{\rm DM}, \psi^{3}_{\rm SM})
\,,
\end{equation}
where $\psi^{3}_{\rm SM}$ generically denotes 
SM third-generation fermions, $\Lambda$ the 
effective scale of new physics, and $\mathcal{O}_i^{(d)}$ the local operators
of dimension $d$.
The $Z_2$ symmetry implies that the operators 
in $\mathcal{L}^{(0)}_\text{SM-DM}$ are bilinear in the DM field. We also require the operators to be bilinear in $\psi^{3}_{\rm SM}$ and to involve no light SM fermions. As a result of Lorentz and gauge invariance, the operators with lowest canonical dimension in 
$\mathcal{L}^{(0)}_\text{SM-DM} $ have $d=6$.

We are interested to analyse the potentially strongest constraints on the effective scale of the interactions from  direct-detection experiments 
in absence of fine-tuned cancellation mechanisms.
To this purpose,  we focus on limits derived by the  spin-independent (SI) cross-section, which are stronger than spin-dependent ones due to coherent nuclear enhancement~\cite{Agrawal:2010fh}.

Since direct-detection experiments probe interactions between DM and nucleons, the first step is to identify the effective couplings between DM and light quarks ($u$, $d$), which arise at the one-loop level.
Among all the possible DM-light quark operators induced, only a subset contributes significantly to SI scattering, as most others are suppressed by the small momentum transfer characteristic of direct detection ($|\mathbf{q}| \ll m_{\rm nucleon}$), see, e.g., Ref.~\cite{Cirelli:2013ufw}. The leading operators relevant for SI interactions in the fermionic case ($\psi_{\rm DM}= \chi$) are
\begin{align}
\mathcal{O}_{VV}^{\chi q}&=(\bar{\chi} \gamma^\mu \chi)(\bar{q} \gamma_\mu q), \quad\:\:\: \mathcal{O}_{SS}^{\chi q}= m_q\,(\bar \chi  \chi)(\bar{q} q)\,,
\end{align}
while for scalar DM ($\psi_{\rm DM} = \phi$),
the dominant operators are 
\begin{align}
\mathcal{O}_{VV}^{\phi q}&=(\phi^* i \overleftrightarrow{\partial_\mu} \phi)(\bar{q} \gamma_\mu q), \quad \mathcal{O}_{SS}^{\phi q}= m_q\,(\phi^* \phi)(\bar{q} q)\,,    
\end{align}
where $q = u,d$. For both DM candidates,  scalar operators arise from interactions involving the Higgs field, 
hence are strongly suppressed by the light quark mass compared to the vector ones. 

\begin{figure}[t]
\includegraphics[width = 0.3\textwidth]{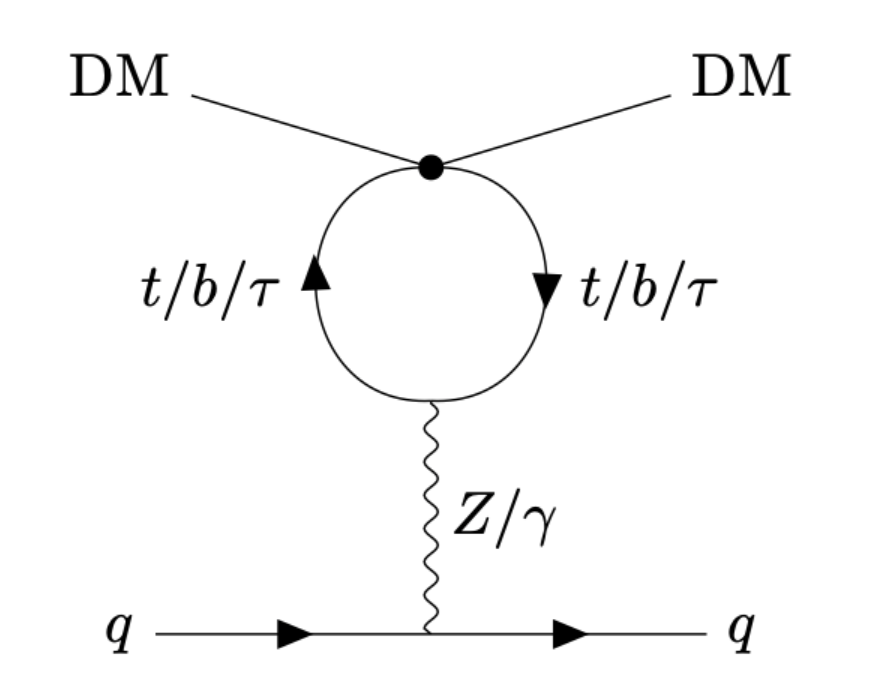} 
    \caption{One-loop Feynman diagram responsible for DM-light quark coupling in our EFT (for both fermion and scalar DM case).
    The dot represents the generic effective operators involving only third-generation fermions in $\mathcal{L}^{(0)}_\text{SM-DM}$. }
     \label{fig:feyn}
\end{figure} 

We thus restrict our analysis to DM-third generation fermions operators that radiatively induce the 
$\mathcal{O}_{VV}$ light-quark operators. At the one-loop level, these arise via the electroweak diagrams shown in Fig.~\ref{fig:feyn}. Evaluating these diagrams we can translate the direct-detection  bounds into constraints on the high-scale effective operators in $\mathcal{L}^{(0)}_\text{SM-DM}$.

\subsection{Effective Operators for fermion Dark Matter}\label{sec:matchingFermionic}

In the fermionic-DM case, the set of dimension-6 effective operators that generate the leading DM-light quark interactions at the one-loop level are 
\begin{align}
    \mathcal{O}_{VL}^{\chi Q} &= (\bar \chi \gamma_\mu \chi) (\bar Q^3 \gamma^\mu Q^3)\,, \nonumber\\
    \mathcal{O}_{VR}^{\chi t_R} &= (\bar \chi \gamma_\mu \chi) (\bar t_R \gamma^\mu t_R)\,, \nonumber\\
    \mathcal{O}_{VR}^{\chi b_R} &= (\bar \chi \gamma_\mu \chi) (\bar b_R \gamma^\mu b_R)\,, \label{eq:opferm} \\
    \mathcal{O}_{VL}^{\chi L} &= (\bar \chi \gamma_\mu \chi) (\bar L^3 \gamma^\mu L^3)\,, \nonumber\\
    \mathcal{O}_{VR}^{\chi \tau_R} &= (\bar \chi \gamma_\mu \chi) (\bar \tau_R \gamma^\mu \tau_R)\,. \nonumber
\end{align}
Here $ Q^3, L^3$ denote the third generation SM $SU(2)_L$ doublets, and $t_R, b_R,\tau_R$ the quark and lepton weak singlets. 

In principle, tensor operators could also generate vector interactions with light quarks at the one-loop level; however, tensor operators in 
$\mathcal{L}^{(0)}_\text{SM-DM}$ starts at $d=7$ and are expected to be suppressed in the limit $\Lambda \gg m_\chi$. This is why we do not include them in our analysis.  It is nevertheless worth noting that  tensor operators lead to a low-energy matrix-element for nucleon-DM proportional to $m_\chi$, which can  become significant for $m_\chi \sim \Lambda$. On the other hand, scalar operators are completely negligible starting at $d=7$  and yielding light-quark matrix-elements suppressed by small Yukawa couplings. 

For simplicity, in the following we denote the Wilson coefficients of the operators in \eq{eq:opferm}
as $C_{VX}^{\chi f}$, where $X$ can be either $L$ with $f= t,b,\tau,\nu_\tau$ or $X=R$ with  $f= t,b,\tau$. With this notation,
we have to enforce the high-scale condition  $C_{VL}^{\chi t}=C_{VL}^{\chi b} = C_{VL}^{\chi Q}/2$
and $C_{VL}^{\chi \tau}=C_{VL}^{\chi \nu_\tau}=  C_{VL}^{\chi L}/2$ to recover the five independent terms in \eq{eq:opferm}. 

The one-loop matrix element for DM scattering to light quarks, via both $Z$ and $\gamma$ exchange, 
can be decomposed as
\begin{equation}
\label{eq:Cq}
  \mathcal{M} = -i\, C_{VV}^{\chi q} \left( \bar \chi \gamma^\mu \chi \right)  \left( \bar q \gamma_\mu q \right) 
    - i\, C_{VA}^{\chi q} \left( \bar \chi \gamma^\mu \chi \right) \left( \bar q \gamma_\mu \gamma_5 q \right)\,,
\end{equation}
where  $q=u,\, d$.  The vector-axial operator, which is responsible for spin-dependent scattering, will be neglected in the following,
while the leading $C_{VV}^{\chi q}$ coefficient, evaluated in the zero recoil limit is 
\begin{align}
   C_{VV}^{\chi q} =& -\sum_{f=t,b,\tau}N_c^f\bigg[\dfrac{ g^2 \, g_A^f\, g_V^q \,m_f^2}{4 \pi^2 \cos^2(\theta_W) M_Z^2 }\left(C_{VL}^{\chi f}-C_{VR}^{\chi f}\right) \nonumber\\
  & + \dfrac{  e^2\,Q_f \,Q_q }{24 \pi^2} \left(C_{VL}^{\chi f}+C_{VR}^{\chi f}\right)\bigg]
   \log\left(\dfrac{\mu_{\rm UV}^2}{m_f^2}\right)\,.
   \label{CVVferm}
\end{align} 
with
\begin{equation}
g_{V}^{q} = (1/2)T^q_3 - Q_{q} \sin^2\theta_W\,, \quad g_{A}^{f} = (1/2)T^f_3\,.
\end{equation}
Here  $N_c^f$, $Q_{f}$, and  $T^f_3$ denote, respectively, number of colours,  electric charge, and third component of  weak isospin for the fermion $f$, while $\theta_W$ denotes the weak mixing angle. The ultraviolet (UV) divergent part of the matrix element 
has been removed in the $\overline{\mathrm{MS}}$ scheme, leading to the appearance 
of the (high-scale) renormalization scale $\mu_{\rm UV} \sim \Lambda$.
 
In principle, one can obtain a renormalization-group improved expression of $C_{VV}^{\chi q}$ re-summing the logarithms generated between the high scale $\Lambda$ and a low scale below $m_f$. However, for $\Lambda \sim 1$~TeV the QED-induced logarithms are not large.
Moreover,  the effective running range varies for the different amplitudes, with the leading $Z$-penguin contribution being free from large logs and relevant only for the top quark. A further potential improvement is the inclusion of two-loop QCD-induced running; however, this turns out negligible but for the $b$-quark case (where it does not exceed the one-loop induced electroweak amplitude). 
In absence of a complete UV  theory, all these effects can be reabsorbed by $O(1)$ redefinitions of the initial values of the Wilson coefficients, which are beyond our control.  This is why, for simplicity, we keep the one-loop expression in (\ref{CVVferm}) setting $\mu_{\rm UV}=1$~TeV.

\subsection{Effective Operators for scalar Dark Matter}\label{matchingScal}
In the case of scalar DM, the leading high-scale operators are 
\begin{align}
\mathcal{O}^{\phi Q}_L &=(\phi^* i \overleftrightarrow{\partial_\mu} \phi)\left(\bar Q^3\gamma^\mu Q^3\right)\,, \nonumber \\
\mathcal{O}^{\phi t}_R &=(\phi^* i \overleftrightarrow{\partial_\mu} \phi)\left(\bar t_R \gamma^\mu t_R\right)\,,   \nonumber \\
\mathcal{O}^{\phi b}_R &=(\phi^* i \overleftrightarrow{\partial_\mu} \phi)\left(\bar b_R \gamma^\mu b_R\right)\,, \\
\mathcal{O}^{\phi L}_L &=(\phi^* i \overleftrightarrow{\partial_\mu} \phi)\left(\bar L^3 \gamma^\mu  L^3\right)\,, \nonumber\\
\mathcal{O}^{\phi \tau}_R &=(\phi^* i \overleftrightarrow{\partial_\mu} \phi)\left(\bar \tau_R \gamma^\mu \tau_R\right)  \,.  \nonumber
\end{align}
Proceeding as in the fermionic case, the coefficient of the vector operator involving light quarks in 
zero recoil limit is 
\begin{align}
   C_{VV}^{\phi q} =& -\sum_{f=t,b,\tau}N_c^f\bigg[\dfrac{ g^2 \, g_A^f\, g_V^q \,m_f^2}{4 \pi^2 \cos^2(\theta_W) M_Z^2 }\left(C_{L}^{\phi f}-C_{R}^{\phi f}\right) \nonumber\\
 &  + \dfrac{  e^2\,Q_f \,Q_q }{24 \pi^2} \left(C_{L}^{\phi f}+C_{R}^{\phi f}\right)\bigg]
   \log\left(\dfrac{\mu_{\rm UV}^2}{m_f^2}\right)\,.\label{CVVscalfull}
\end{align}

\section{Direct detection bounds}
\label{sec:DDbounds}
Using the matching described in \sect{sec:matchingFermionic}, we recast the experimental direct-detection  bounds 
into constraints on the high-scale coefficients of the EFT involving only third-generation fermions. Specifically, for the fermionic DM case, we build upon the results presented by the XENON1T collaboration in~\cite{XENON:2022avm}, where constraints were derived for light-quarks Wilson coefficients $ C_{VV}^{\chi q} $ defined in \eq{eq:Cq}. To refine these limits, we incorporate the latest constraints on the spin-independent DM-nucleon cross section from the LUX-ZEPLIN collaboration~\cite{LZ:2024zvo}.  Concretely, we rescale the bounds on $ C_{VV(A)}^{\chi q} $ obtained by XENON1T, which were based on their now-superseded spin-independent DM-nucleon cross section, using the updated cross-section bounds from LUX-ZEPLIN.

\begin{figure}[t]
\center
\includegraphics[width = 0.5\textwidth]{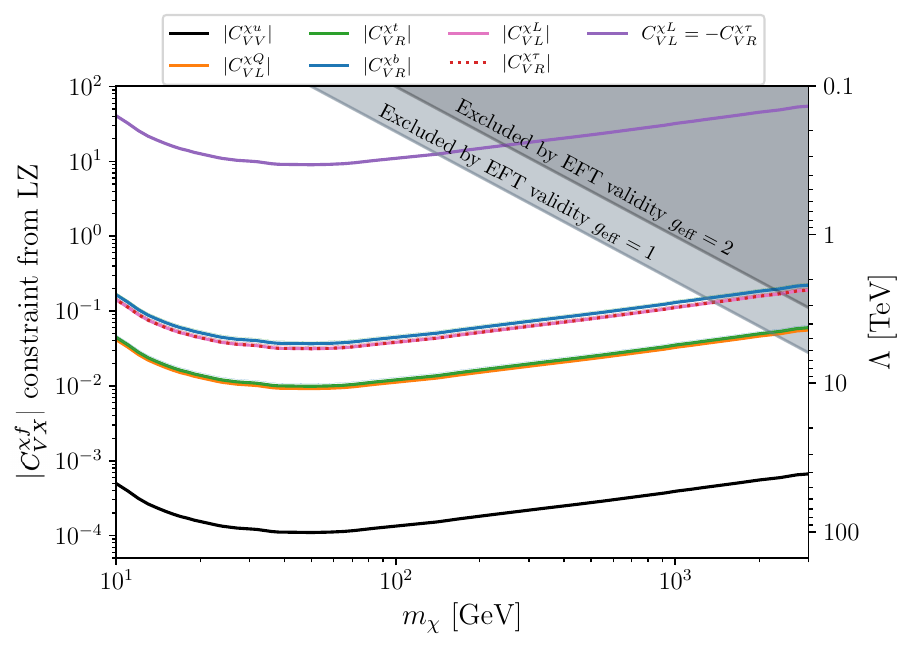} 
    \caption{Constraints on the Wilson coefficients of the EFT at the high-scale in the fermionic case (colored lines). On the left axis we assume $\Lambda=1$ TeV, while on the right axis we display the NP scale $\Lambda$, assuming $C^{\chi f}_{VX}=1$.  For comparison, the case of 
    an operator with unsuppressed coupling to light quarks, not present in the EFT, is also shown (black line). 
    The gray regions are excluded by the breaking of the EFT validity (see text).
\label{Fig:Matching_Fermionic_CVVu}
     }
\end{figure}

In Figure~\ref{Fig:Matching_Fermionic_CVVu}, we present the bounds on the various Wilson coefficients considered for the fermionic DM case. We analyse both the case of a single operator at a time and a scenario where we impose the relation \( C_{VL}^{\chi L} = -C_{VR}^{\chi\tau} \) as an illustrative case. This specific choice cancels the photon contribution in Eq.~\eqref{CVVferm}, effectively relaxing the bound by two orders of magnitude.  

Remarkably, direct-detection constraints fall in the few-TeV range,  which is the natural expectation of this EFT,  
given also the constrains from collider experiments~\cite{Allwicher:2023shc}. This implies that direct-detection 
experiments have just started to probe the most interesting parameter range for third-generation-philic dark matter.
Among the operators considered, \( \mathcal{O}_{VL}^{\chi Q} \) is the most constrained; however, also in that case the bound on the effective scale does not exceed  10~TeV (but for a narrow region around $m_\chi \sim 50$~GeV). 
Conversely, the combination \( C_{VL}^{\chi L} = -C_{VR}^{\chi\tau} \) remains the least constrained, with \( \Lambda \gtrsim 0.3 \) TeV.  
For reference, the bounds on operators coupling directly (not loop-suppressed) DM to light quarks
set  constrains on $\Lambda$  from about 30 TeV to 100 TeV, well above the range dictated by the electroweak hierarchy problem. 
We can thus conclude that the hypothesis of third-generation-philic WIMP can save, at least with current bounds, the basic idea of a connection between NP responsible for DM and that responsible for the stabilisation of the electroweak  scale.

For DM masses approaching the scale \( \Lambda \), the validity of the EFT framework must be carefully assessed. 
Assuming the operators in $\mathcal{L}^{(0)}_\text{SM-DM}$ are obtained by the tree-level exchange of a mediator 
of  mass $M_{\rm med}$ and effective coupling $g_{\rm eff}$ (to both DM and SM fermions), we expect 
\begin{equation}
\frac{C^{(6)}_i }{\Lambda^2} \approx  \frac{ g_{\rm eff}^2 }{ M_{\rm med}^2 }\,.
\label{eq:Mmed}
\end{equation}
Based on kinematical considerations for DM annihilation, we assume a breakdown of the EFT description if 
$m_\chi > M_{\rm med}/2$, with  $M_{\rm med}$ extracted inverting \eq{eq:Mmed}. 
In \fig{Fig:Matching_Fermionic_CVVu} we indicate this region for two reference choices of the effective 
coupling: \( g_{\rm eff} = 1 \) and \( g_{\rm eff} = 2 \).

For the scalar DM case, we follow a similar approach. We consider the bounds on individual effective operators coupling DM to light quarks, as reported in~\cite{Cirelli:2013ufw}, and rescale them using the latest constraints on the spin-independent DM-nucleon cross section from the LUX-ZEPLIN collaboration. 
We then use the matching reported in \sect{matchingScal} to recast the bounds on the operators involving third generation SM fermions.
Since the matching of the fermionic and scalar DM-third generation operators onto non-relativistic operators is identical, and their contributions to SI interactions are also the same (see Eqs.~\eqref{CVVferm} and \eqref{CVVscalfull}), the resulting bounds for the scalar DM candidate are effectively identical to those in the fermionic case. Therefore, we do not display them here explicitly.

\section{Relic abundance}
\label{sec:relics}

The operators discussed in \sect{sec:setup} contribute to thermal DM production in the early universe. It is crucial to determine whether the DM abundance generated through these channels is consistent with both the observed DM density and the constraints derived from direct detection searches.

\begin{figure}[p]
\center
\includegraphics[width = 0.4\textwidth]{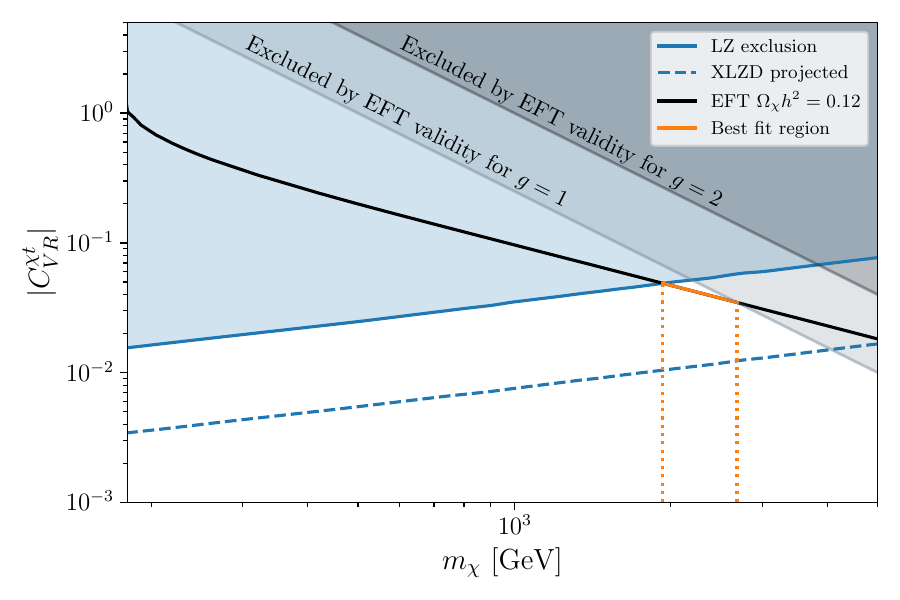} 
\includegraphics[width = 0.4\textwidth]{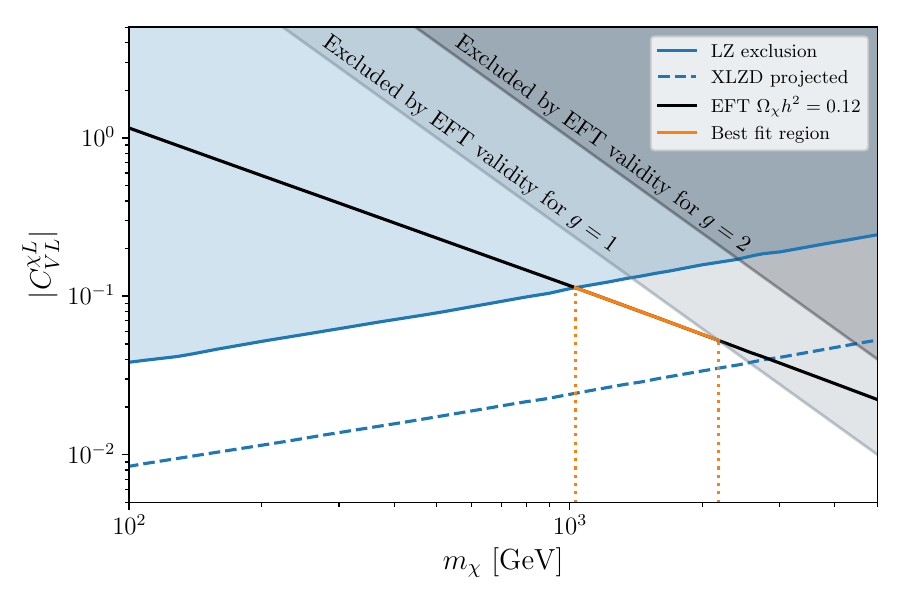} 
\includegraphics[width = 0.4\textwidth]{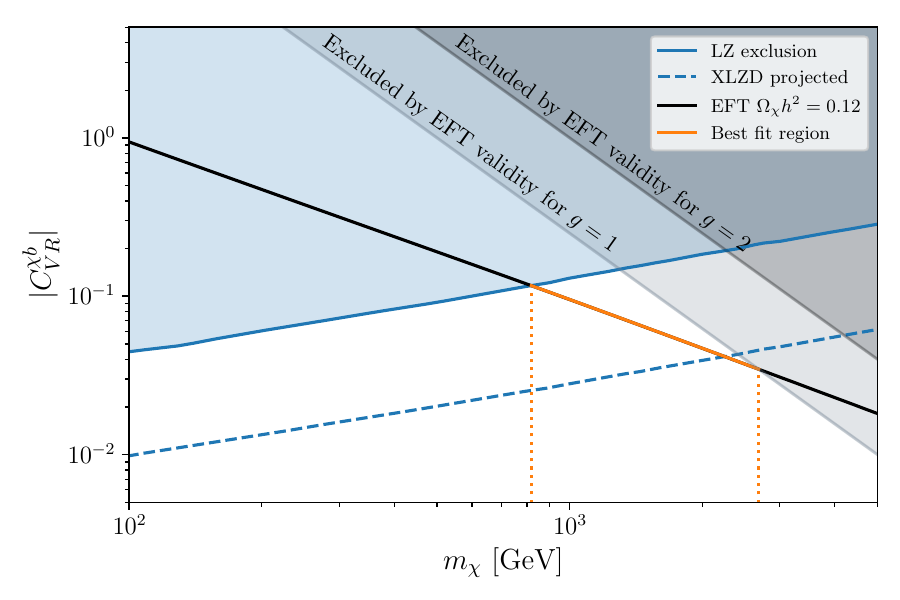}  
\includegraphics[width = 0.4\textwidth]{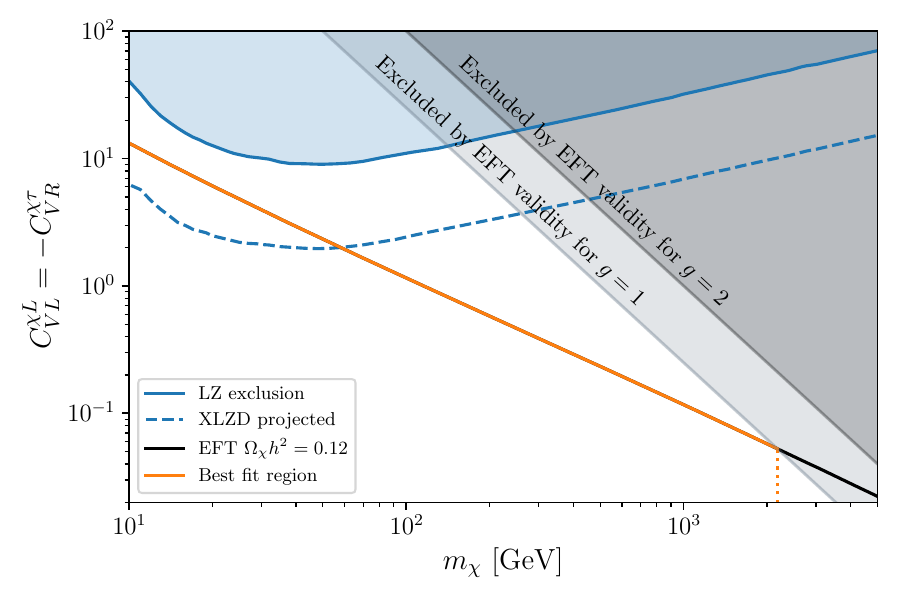} 
    \caption{\label{Fig:BoundsFermionicDMEFT}
    Constraints on the Wilson coefficients of four representative effective operators in the fermionic-DM case,  for $\Lambda=1$ TeV.
    The blue region is the currently excluded region by LUX-ZEPLIN direct searches, the blue dashed line indicates the future sensitivity of XLZD,  the grey regions are excluded by the EFT validity (see text). The black/orange line corresponds to the condition $\Omega_{\rm \chi}h^2=0.12$, where the  DM candidate saturates the observed relic abundance (within the EFT),
  with the orange part highlighting the viable mass range considering all constraints.}
\end{figure}

\begin{figure}[t]
\center
\includegraphics[width = 0.4\textwidth]{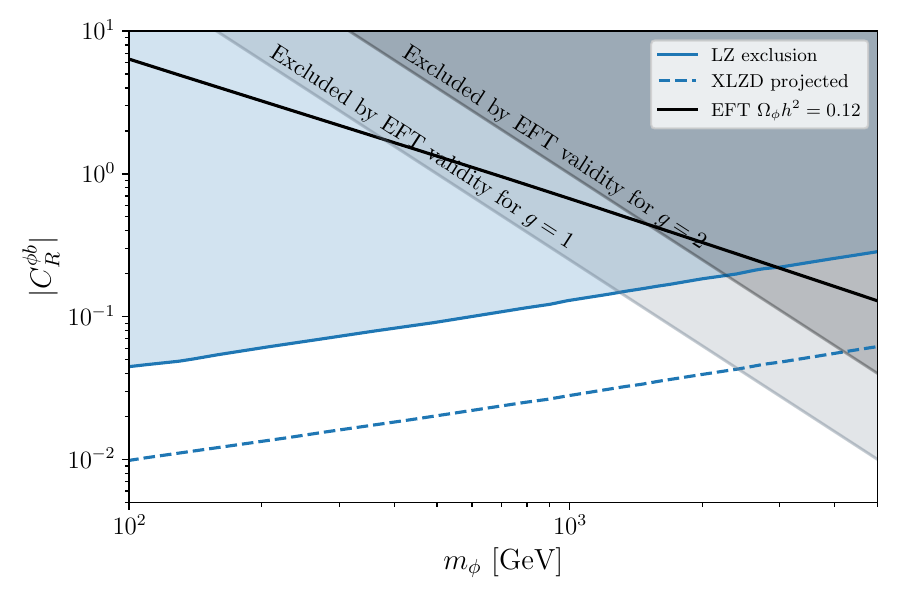}  
\includegraphics[width = 0.4\textwidth]{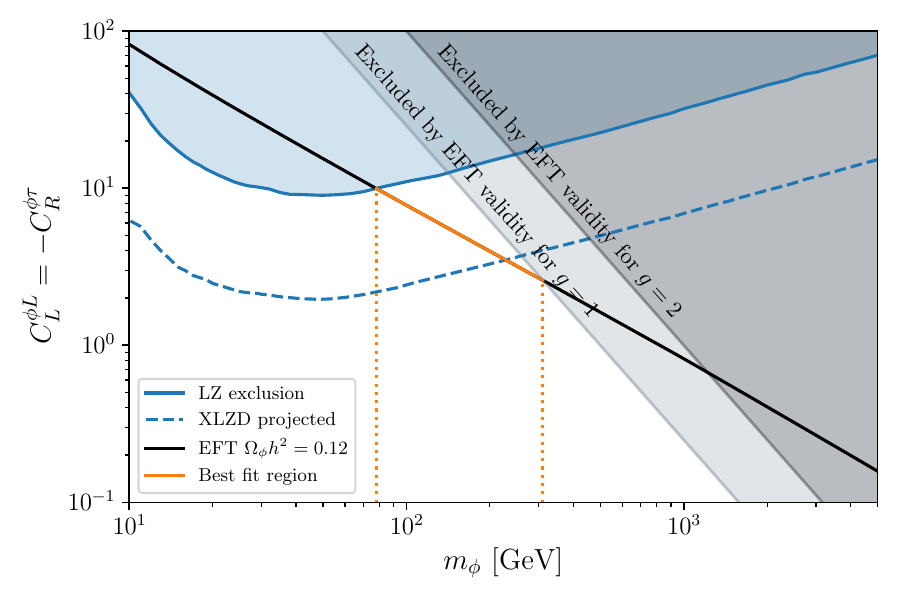} 
    \caption{\label{Fig:BoundsScalarDMEFT}
    Constraints on the Wilson coefficients of two representative effective operators in the scalar-DM case,  for $\Lambda=1$ TeV.
    Notations as in Fig.~\ref{Fig:BoundsFermionicDMEFT}.}
\end{figure}

To this end, we computed the thermally averaged cross section for the process $ \mathrm{DM\:DM} \to \bar{f} f $, where $ f $ is a third-generation SM fermion, mediated by the effective operators introduced in \sect{sec:setup}. This process dictates the DM annihilation rate and, consequently, the final relic abundance. The evolution of the DM number density is governed by the Boltzmann equation, whose solution determines whether the produced DM abundance aligns with the observed cosmological value or whether additional mechanisms need to be considered. The results are presented in~\fig{Fig:BoundsFermionicDMEFT}  and 
\fig{Fig:BoundsScalarDMEFT} for representative fermionic- and scalar-DM operators, respectively. The black/orange lines denote the regions of parameter space that yield the observed DM relic density, $\Omega_{\rm DM} h^2 = 0.12$, where $\Omega_{\rm DM}$ is the DM density fraction relative to the critical density of the universe, and $h$ is the dimensionless Hubble parameter defined 
as $h = H_0 / (100 \, \text{km} \, \text{s}^{-1} \text{Mpc}^{-1})$. 

Both in~\fig{Fig:BoundsFermionicDMEFT} and in~\fig{Fig:BoundsScalarDMEFT} 
 the orange region highlights the range for the DM mass where a given EFT operator account for the entire relic abundance, while being in agreement with both direct-detection and EFT validity bounds. In the fermion case, considering a single operator at a time, an allowed region with $m_\chi$ between 1~and 2~TeV is always found. The smaller and wider regions being associated to
$C_{V L}^{\chi t}$ and $C_{V R}^{\chi b}$, respectively (the cases not explicitly shown in \fig{Fig:BoundsFermionicDMEFT} falls within these two ranges). Hence present data leave open the parameter space where the DM candidate is close in mass to the mediator responsible its interaction to  third-generation SM fermions. This is an interesting region from the model-building point of view, if one  assumes a common origin for $m_\chi$  and $M_{\rm med}$. From a phenomenological point of view, it is also very interesting to note that this region will be almost entirely probed by the   projected sensitivity of XLZD~\cite{XLZD:2024nsu}, that should improve the current bounds by almost one order of magnitude.  The only case where this does not happen is in the fine-tuned limit $C_{VL}^{\chi L}=-C_{VR}^{\chi \tau}$, where 
direct-detection constraints are much weaker and  $m_\chi$ is essentially unconstrained. 

In contrast, in the scalar DM scenario, no viable parameter space survives if one considers only one operator at a time: the region compatible with direct-detection constraints would result in an overproduction of DM, rendering such scenario cosmologically inconsistent. This can be circumvented only in the fine-tuned limit  $C_{L}^{\phi L}=-C_R^{\phi \tau}$, where a viable parameter space 
for $m_\phi~O(100)$~GeV emerges.  

\begin{figure}[p]
\center
\includegraphics[width = 0.4\textwidth]{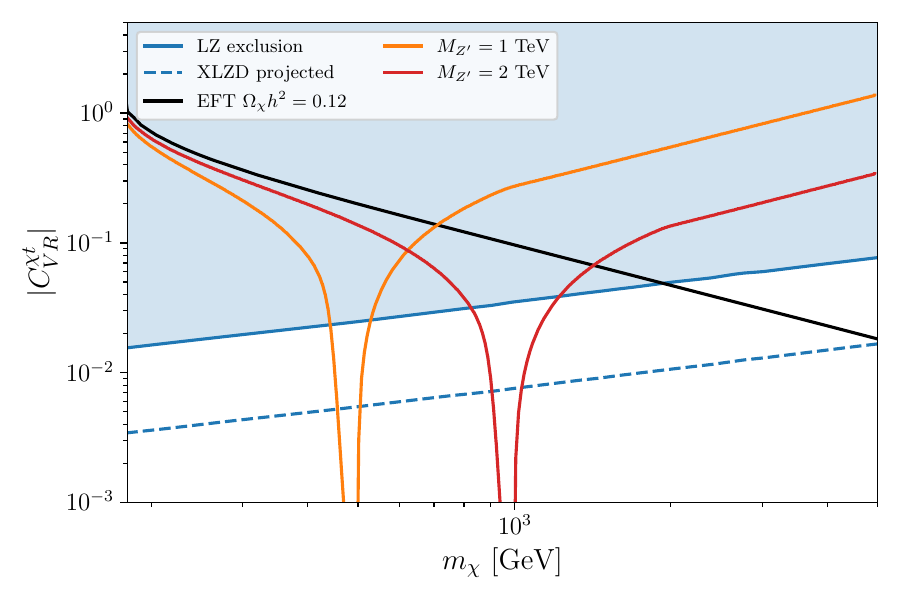} 
\includegraphics[width = 0.4\textwidth]{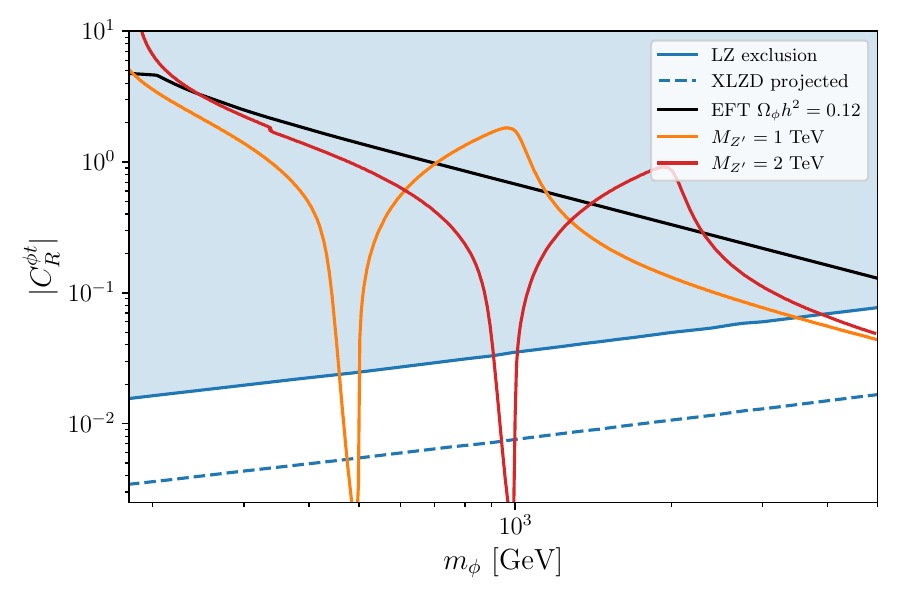}
\includegraphics[width = 0.4\textwidth]{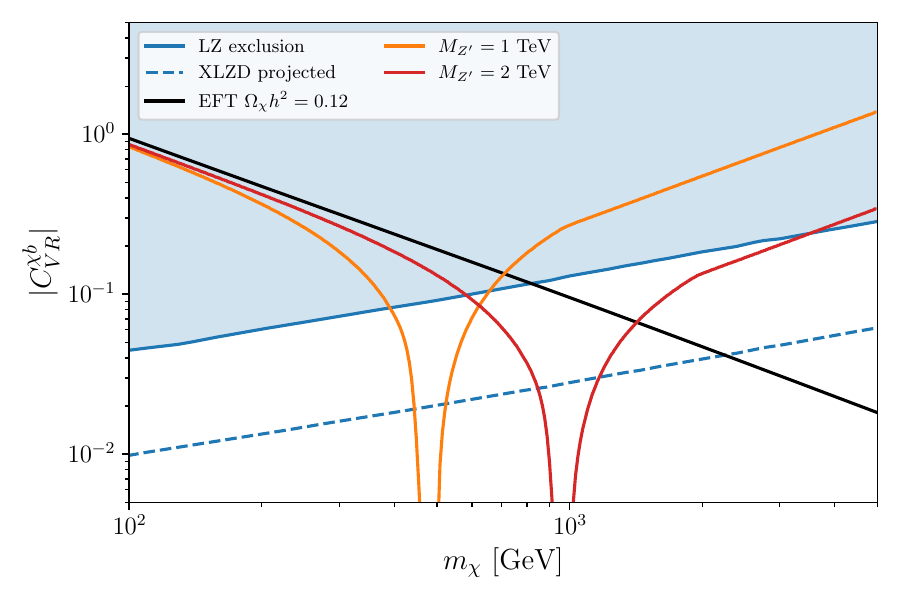} 
\includegraphics[width = 0.4\textwidth]{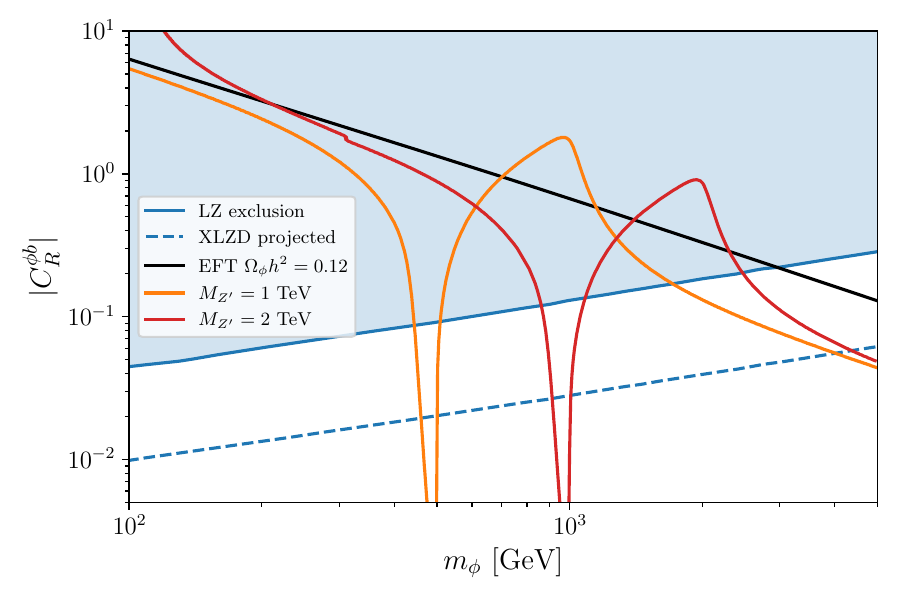} 
    \caption{\label{Fig:BoundsFermionicDMZp} 
    Bounds on the Wilson coefficients 
    of two representative effective operators in the fermionic and scalar DM case, for $\Lambda=1$~TeV.  The blue region is excluded by LUX-ZEPLIN DM direct searches, the blue dashed line indicates the future sensitivity of XLZD. The black line corresponds to $\Omega_{\rm \chi}h^2=0.12$ in the EFT case ($\Lambda=1$ TeV), while the orange and red lines correspond to $\Omega_{\rm \chi}h^2=0.12$ for a mediating vector boson with mass $M_{Z^\prime}=1$~TeV and 2~TeV, respectively.}
    \end{figure}

\subsection{Beyond the EFT approach: resonant DM production}

The EFT approach is always well suited to analyse
direct-detection bounds, which  involve low energies;
however, it might miss relevant features when analysing the relic-abundance constraint, which involves high-energy processes. More precisely, the regions excluded in~\fig{Fig:BoundsFermionicDMEFT} and \ref{Fig:BoundsScalarDMEFT} due to a too small  
DM  annihilation cross section, within the EFT, can be rescued in specific UV models with light mediators and/or additional states enabling DM co-annihilation. 
We illustrate this point here in a simple case, allowing for DM resonant production and annihilation. 

If the $\mathrm{DM\:DM} \to \bar{f} f $ process happens in the $s$ channel and the DM mass approaches half of the mediator mass, the process is resonantly enhanced, altering the conclusions derived in the EFT approach. To explore how this possibility could modify the results in \fig{Fig:BoundsFermionicDMEFT} and \ref{Fig:BoundsScalarDMEFT}, we assume that the effective operators are generated by an appropriate $Z^\prime$ vector boson~\cite{Langacker:2008yv} that mediates the interactions between the DM candidate and the third-generation SM fermions.\footnote{For simplicity we assume that the new \( Z^\prime \) boson does not mix with the SM \( Z \) boson.} Such a flavor non-universal $Z^\prime$ is a natural expectation in many models based on the hypothesis of flavor deconstruction~\cite{Davighi:2023iks,FernandezNavarro:2023rhv,Davighi:2023evx,Barbieri:2023qpf,Greljo:2024ovt} 
(see also~\cite{Andrianov:1998hx,Altmannshofer:2016jzy,Correia:2016xcs,Correia:2019pnn,Correia:2019woz}).
We implemented this scenario in micrOMEGAs \cite{Alguero:2023zol}, consistently including the full dependence on the  $Z^\prime$ mass from the 
propagator, rather than relying on the EFT limit. We then computed the relic DM abundance using micrOMEGAs. 

In \fig{Fig:BoundsFermionicDMZp}, we compare the results obtained within the EFT framework to those from the full $Z^\prime$ exchange calculation, shown for two benchmark $Z^\prime$ masses, for both fermionic and scalar DM.
As expected, the EFT approximation and the full theory coincide for small DM masses, where the mediator remains far off-shell and its effects can be effectively captured by local operators. However, near the resonant region, where $ m_{\rm DM} \approx M_{Z^\prime}/2 $, the two approaches yield significantly different results. The inclusion of the $Z^\prime$ exhibits the expected resonance effect, which leads to a sharp drop in the DM abundance due to a stronger annihilation cross section. 
With a suitable choice of the $Z^\prime$ mass one can always find a parameter region consistent with the observed relic abundance and the direct-detection constraints. Of course, this requires some amount of fine tuning, but it is not such a difficult condition to realize, especially if the DM candidate and the mediator have a common origin. 
The inclusion of an explicit mediator serves as a rudimentary example illustrating how a more complete UV framework — featuring additional DM annihilation channels — can open up viable regions of parameter space that may be accessible to future experiments such as XLZD.

\section{Conclusions}  
\label{sec:concl}

The hypothesis of a multi scale UV completion of the SM, where the first energy threshold above the electroweak scale hosts dynamics coupled mainly to the third-generation of SM fermions, is both theoretically appealing and of great  phenomenological interest. In this work, we have investigated how  dark matter could fit within this general picture from a general bottom-up EFT perspective.  To do so, we have investigated the implications of effective operators coupling dark matter to third-generation SM fermions, both for fermionic and scalar DM candidates, analysing the corresponding bounds from direct detection and relic abundance constraints.

From the direct-detection perspective, we have recasted the latest LUX-ZEPLIN limits onto DM-third-generation vector couplings using one-loop matching to light quark currents. Although loop suppression weakens the direct-detection reach, it is precisely this suppression that enables an important shift in perspective: if dark matter couples predominantly to third-generation fermions, then the scale of new physics can naturally lie in the few TeV region.  This stands in contrast to scenarios where dark matter couples directly to light quarks, which are subject to stringent bounds pushing the new physics scale up to $\sim 100$~TeV. Thus, the assumption of DM coupled mainly to the third-generation provides a viable path to lower-scale new physics, within reach of current and future experiments. Most importantly, this hypothesis, which is theoretically well motivated, leaves open the possibility of a connection between the DM problem and the electroweak hierarchy problem, as in the original spirit of the WIMP paradigm. 

From the relic-abundance perspective, we have shown that imposing the observed relic abundance from thermal freeze-out, within the EFT approach, leads to strong restrictions of the allowed parameter space.
In particular,  nearly all viable parameter space for scalar DM candidates is excluded. In contrast, a well-defined region for a fermionic DM candidate with mass in the 1–2 TeV range emerges. This is an interesting region from the model-building point of view, which might suggest a common origin for the masses of the DM candidate and that of the corresponding mediator(s). 
We have also shown that this region will be fully probed by upcoming direct-detection experiments. 

When analysing the relic-abundance constraint, we have also illustrated the limitation of the EFT approach.  While the EFT description is valid and accurate for low DM masses, it fails to capture resonant effects and new annihilation channels that arise when the DM mass approaches the mediator mass. By incorporating a simple UV completion with a vector mediator, we showed that the resonant production can substantially deplete the DM relic density, modifying the viable parameter space as compared to the EFT result. 

Overall, our study demonstrates that direct-detection experiments are currently  probing the most interesting parameter region of a general class of well-motivated models, based on the hypothesis of a TeV-scale DM candidate coupled mainly to the third generation. While a large fraction of the parameter space is already ruled out, a significant  portion is still allowed. Interestingly enough, this lies within the reach of the next generation of direct-detection experiments. Our results also underscore the importance of complementing the EFT framework with simplified models when interpreting experimental data, particularly in regions near the breakdown of the effective description.

\subsection*{Acknowledgements}

We thank Laura Baudis, Bj\"orn Penning, and several members of their experimental groups at UZH for interesting discussions on this topic. We are also particularly grateful to Nicolas Angelides, Chiara Capelli, and Davide Racco for organising the lively seminar series where these discussions have originated. This project has received funding from the Swiss National Science Foundation~(SNF) under contract~200020\_204428.

\bibliography{references}
\end{document}